\documentclass[12pt]{article}
\date{}

\begin{document}

\title{The interplay of superconductivity and localization in
Nd$_{2-x}$Ce$_x$CuO$_{4+\delta}$ single crystal films}


\author{G.\,I. Harus, A.\,I. Ponomarev\footnote{e-mail: ponomarev@imp.uran.ru},
T.\,B. Charikova, A.\,N. Ignatenkov,\\
L.\,D. Sabirzjanova, N.\,G. Shelushinina,\\
{\normalsize\it Institute of  Metal Physics, RAS, 620219 Ekaterinburg, Russia}\\
V.\,F. Elesin, A.\,A. Ivanov, I.\,A. Rudnev\\ {\normalsize\it Moscow Engineering
Physics Institute, 115410 Moscow, Russia}}


\maketitle

\abstract{The influence of nonstoichiometric disorder on the in-plane resistivity
and SC-transition has been investigated for Nd$_{2-x}$Ce$_x$CuO$_{4+\delta}$ single
crystal films ($x=0.15$ and 0.18). It is shown that with increasing of $\delta$ the
in-plane normal state resistivity increases (the mean free path diminishes) and
SC-transition temperature decreases with essential broadening of the transition
region. The observed evolution from homogeneous metallic (and superconducting)
Nd$_{2-x}$Ce$_x$CuO$_{4+\delta}$ system to inhomogeneous dielectric one is described
as Anderson-type disorder-induced transition in a two-dimensional electron system.}


\section{Introduction}

Single crystal Nd$_{2-x}$Ce$_x$CuO$_4$ belongs to a class of basically tetragonal copper oxide
compounds. All of these crystals have CuO$_2$ planes in their structure separated by buffer
layers of other atoms. The conduction process mainly occurs in the CuO$_2$ planes while the
other layers act as so called charge reservoirs simply providing carriers [1].

The structure of Nd$_2$CuO$_4$  crystal called T$^{\prime}$ structure is the simplest among
the cuprates: in the T$^{\prime}$ structure the apical oxygen atoms are displaced so as to
make an isolated CuO$_2$ plane ($ab$-plane). Upon doping Nd$^{3+}$ ions are randomly replaced
by Ce$^{4+}$ ions and an excess of electrons is donated to the CuO$_2$ planes resulting in the
$n$-type in-plane conduction. The electrons are concentrated within the confines of conducting
CuO$_2$ layers separated from each other by a distance $c\cong 6$\AA. Strong resistivity
anisotropy ($\rho_c/\rho_{ab}\cong 10^4$) is observed in Nd$_{2-x}$Ce$_x$CuO$_{4+\delta}$
single crystals mainly caused by the two-dimensional (2D) nature of the system. Due to the
highly layered structure Nd$_{2-x}$Ce$_x$CuO$_4$ single crystal may be regarded as a natural
superlattice or more exactly as a selectively doped multi-quantum well system with CuO$_2$
layers acting as wells and buffer NdO layers as Ce - doped barriers [2].

We report here a study of a disorder influence on the in-plane transport and  low-temperature
localization effects in Nd$_{2-x}$Ce$_x$CuO$_{4+\delta}$ single crystal films deposited by
flux separation technique [3]. The different degree of disorder has been got by varying of
annealing conditions or by ion irradiation of a sample. It is of importance that in this model
two-dimensional system there is an opportunity to vary gradually a disorder in the sample and
thus to investigate 2D-localization process in detail. We compare our results with the data of
preceding reports on disorder induced insulator-metal (insulator-superconductor) transition in
Nd$_{2-x}$Ce$_x$CuO$_{4+\delta}$ system with $x=0.18$ [4] and $x=0.15$ [5].

\section{Results and discussion}

The transport properties of Nd$_{2-x}$Ce$_x$CuO$_{4+\delta}$ are extremely sensitive to oxygen
content: an as-grown bulk crystal ($x=0.15$) is not superconducting and a small amount of
oxygen ($\delta \approx 0.01 \div 0.03$) has to be removed in order to achieve
superconductivity [6, 7]. It is ascertained now [5-8] that as-grown
Nd$_{2-x}$Ce$_x$CuO$_{4+\delta}$ crystals have an excess oxygen atoms ($\delta > 0$) at
interstitial out-of-plane sites which random potential substantially disturbs the conductive
CuO$_2$ plane and may cause localization of the electrons provided by Ce doping. The annealing
in vacuum gradually reduces the concentration of the oxygen nonstoichiometric defects and the
optimal reduction gives the composition close to stoichiometric one ($\delta \rightarrow 0$).

In the model of autonomous CuO$_2$ planes the partial 2D - conductivity of a plane may be
obtained as $\sigma_s = \rho_{ab}^{-1}c$. From the relation $\sigma_s = (k_F\ell)e^2/h$
($\ell$-mean free path, $k_F$\ - Fermi wave vector) the value of $k_F\ell$ may be estimated.
The parameter $k_F\ell$ is usually regarded as a measure of disorder in a system [9]. Fig.1
shows temperature dependencies of the in-plane resistivity $\rho_{ab}$ for three samples with
$x=0.18$: unreduced (as-grown) sample 1, optimally reduced in vacuum (at 800$^{\circ}$C during
40\,min.) sample 3 and intermediately reduced sample 2. With varying of annealing procedure in
our Nd$_{1.82}$Ce$_{0.18}$CuO$_{4+\delta}$ system the disorder parameter $k_F\ell$ changes in
an order of magnitude from one sample to another (see Fig.1a). Decreasing disorder we first
observe superconductivity for sample 2 with $k_F\ell = 2.5$ in rather good accordance with the
data of Tanda et al. for Nd$_{1.82}$Ce$_{0.18}$CuO$_{4+\delta}$ single crystal epitaxial films
[4]. In their work the low-temperature superconductor-insulator transition tuned by disorder
introduced at various stages of oxygen reduction was fixed at $\rho_{ab}$($T\,=\,18$\,K) =
0.5\,m$\Omega$\,cm, i.e. for $k_F\ell \cong 3$.

Fig.1b demonstrates the low-temperature up-turn of resistivity owing to localization of
carriers. It will be seen that the effect of localization becomes less pronounced with the
decrease of disorder (increase of $k_F\ell$): in sample 3 with $k_F\ell=25$ only marginal
effect of weak localization is observed at $T<4$\,K in magnetic field of $1.3$\,T [10].

In Figs.2 the results of magnetoresistance measurements for sample 2 are presented at magnetic
fields $B$ both parallel and perpendicular to $ab$-plane. As seen the parallel field up to
5.0\,T suppressing superconductivity does not influence on the localization up-turn of normal
state resistivity at $14 < T < 20$\,K. In contrast to it the perpendicular field of 2\,T to a
great extent suppresses localization effect (so as superconductivity). Such behaviour is
typical just for 2D system with weak localization as only magnetic field perpendicular to
plane may destruct an interference of electron diffusion loops leading to localization.

The in-plane resistivity data for five Nd$_{2-x}$Ce$_x$CuO$_{4+\delta}$ single crystals with
$x = 0.15$ are shown in Fig.3. These films of thickness t = 2000\AA\ are annealed at various
conditions: sample 2 is as-prepared, sample 1 is oxygenated (1\,atm., 60\,min.,
780$^{\circ}$\,C). The annealing procedure for sample 5 (10$^{-2}$\,mm\,Hg, 60\,min.,
780$^{\circ}$\,C) is close to optimal one ($T_c=23.5$\,K with $\Delta\,T_c=1.5$\,K) and sample
3, 4 are reduced at intermediate conditions.

Fig.3a shows the decreasing of normal state resistivity value at all temperatures in about of
two orders of magnitude due to process of deoxygenation. The values of parameter $k_F\ell$
estimated from the values of $\rho_{ab}(T=T_{min})$ for samples 1-3 and from the values of
$\rho_{ab}(T=T_c)$ for samples 4 and 5 are presented in Table. The appearance of SC-transition
on the $\rho_{ab}(T)$ dependence with decreasing of disorder in our single crystal films
corresponds to $k_F\ell=10$ (sample 2 with $\rho_{ab}(T_{min}\cong
100$\,K$)=0.15$\,m$\Omega$\,cm, $\rho_{ab}(T_c = 12.5$\,K$)=0.2$\,m$\Omega$\,cm) just as for
bulk single crystal Nd$_{2-x}$Ce$_x$CuO$_{4+\delta}$ with $x = 0.15$ [5] where $\rho_{ab}(T_c
= 11$\,K$)=0.2$\,m$\Omega$\,cm at the border of insulating and superconducting phases.

It should be emphasized that superconducting sample 2 is as-prepared in contrast to the
situation for bulk Nd$_{2-x}$Ce$_x$CuO$_{4+\delta}$ system where as-grown samples are
non-superconducting [5-7] and, in particular, the sample with T$_c=11$\,K in [5] is reduced in
Ar atmosphere at 1000$^{\circ}$\,C for 30\,h in order to induce superconductivity. As it is
argued by Xu et al. [11] in a film besides the normal diffusion channels along $c$-axis and
$ab$-plane additional diffusion channels exist due to grain boundary effects and strains
caused by lattice mismatch between the film and the substrate. Thus it is easier to remove
oxygen in a film than in a bulk crystal of Nd$_{2-x}$Ce$_x$CuO$_{4+\delta}$.

The evolution of low-temperature part of $\rho_{ab}$ dependencies from localization to
superconductivity in tight correlation with $k_F\ell$ value is demonstrated on Fig.3b. A
strong nonmetallic ($d\rho/dT < 0$) dependence at $T \le 100$\,K in sample 1 with $k_F\ell <
1$ and almost completely metallic ($d\rho/dT > 0$) dependencies at all temperature interval $T
> T_c$ in samples 3 - 5 ($k_F\ell \ge 25$) with only slight upturn of $\rho(T)$ with
decreasing of $T$ at $T \le 50$\,K in sample 3 are observed. The most interesting behavior of
$\rho(T)$ is demonstrated by sample 2 which is on the low-temperature border of insulating and
superconducting phases. The logarithmic upturn of $\rho(T)$ in the temperature interval $13\le
T \le 100$\,K takes place (Fig.4a) but the superconductivity comes to light at $T < 13$\,K.

It may be treated as a coexistence of superconductivity and localization or, more exactly,
such a behaviour gives an evidence of the existence of localization effects in the normal
state of material which from the superconductivity develops. It is of importance that the
parameter $k_F\ell > 1$ for sample 2 and thus this normal state is a two-dimensional metal
with weak localization corrections due to quantum interference effects. Magnetic field up to
5.5\,T suppresses superconductivity and reveals localization effects at $T < T_c$ (see the
inset on Fig.4b).

The problem of the mutual interplay of Anderson localization and superconductivity in a
disordered system is extensively discussed in literature (see recent review of Sadovskii
[12]). For 3D Anderson insulator a condition for coexistence of superconductivity and
localization was obtained [13]: the Cooper pairing is possible in localized phase if only
superconducting coherence length (the size of a Copper pair) $\xi$ is much smaller than
localization length: $R_{loc} \gg \xi$.

It is known [9] that in a two-dimensional system with random disorder the radius of electron
localization is of a finite size even in the limit $k_F\ell \gg 1$ (``metallic'' state) and
the following estimate is valid:
\begin{equation}
R_{loc} = \ell\exp{\Bigl(\frac{\pi}{2}\times k_F\ell\Bigr)}. \label{eq:r_1}
\end{equation}
For small mean free path, in the so called ``dirty'' superconductor with $\ell\ll\xi_0$
($\xi_0$ is the coherence length of a pure substance) we have $\xi = (\xi_0\ell)^{1/2}$. As
the disorder diminishes in a 2D-system the coherence length increases with $\ell$ according to
power law while the localization length increases exponentially and thus the condition
$R_{loc} > \xi$ may be attained for $k_F\ell$ of the order of several units. Really, an
analysis of effects of disorder introduced by doping or annealing on electrical properties of
Nd$_{2-x}$Ce$_x$CuO$_{4+\delta}$ and the other copper-oxide superconductors [14] reveals that
low-temperature insulator-superconductor transition is closely related to insulator-metal
transition or, more exactly, to transition from strong to weak localization in a 2D system,
which takes place at $k_F\ell\ge 1$.

Fig.4b shows the dependencies of temperature $T_c$ defined both as the onset of
superconducting transition, $T_c^{onset}$, and as its completion, $T_c(\rho=0)$, on the
parameter $k_F\ell$ for samples 2 - 5 of Nd$_{2-x}$Ce$_x$CuO$_{4+\delta}$ with $x=0.15$. It is
seen that $T_c^{onset}$ gradually diminishes with an increase of disorder but a more vital
effect is the broadening of the transition: for the sample 2, for example, $\Delta T_c \cong
T_c^{onset}$.

Much experimental information has revealed that sufficiently large disorder always suppresses
superconductivity but the ways of $T_c$ degradation may be different. From a theoretical point
of view [12, 15] either amplitude reduction or phase breaking for the pair wave function will
result in suppression of superconductivity. In the case of amplitude reduction, $T_c$
decreases remaining well defined, while in the case of phase breaking $T_c^{onset}$ remains
unchanged but the transition width increases until the material has no region of zero
resistance. It a 2D - system a realization of this two limiting cases depends on the length
scale of a disorder, $R$: either homogeneous or inhomogeneous it is on a scale of the
coherence length $\xi$.

Two idealized classes of 2D - materials: ``homogeneous'' and ``granular'' disordered
superconductors were considered [15 - 17], the ``homogeneous'' system being disordered in an
atomic length scale (e.g. solid solutions) and the ``granular'' system being composed of
regions of relatively clean material (with $R > \xi$) which are separated by regions of normal
conductivity or insulating material. In a ``homogeneous'' system the superconductivity is
believed to be destroyed by a suppression of the pair wave function amplitude while in a
``granular'' system due to increasing fluctuations of its phase. Real systems usually lie in
between these idealized cases.

A typical example of ``homogeneous'' disorder in high-temperature superconductor is, e.g., the
degradation of $T_c$ due to Zn substitution into the CuO$_2$ plane for La$_{2-x}$Sr$_x$CuO$_4$
or YBa$_2$Cu$_3$O$_4$ [18], where a gradual displacement of a well defined SC-transition is
observed. An intriguing examples of ``granular'' superconductors are very thin films of Sn and
Pb [16] where the high resistivity is probably due to weak coupling between small particles of
clean metal evaporated onto glass substrates. With increasing sheet resistance of the films,
the temperature of a sharp onset of superconductivity is apparently not changing but below
$T_c^{onset}$ a long tail with finite resistance develops.

From Fig.4b we see that an oxygenation of Nd$_{2-x}$Ce$_x$CuO$_{4+\delta}$ system leads to
essential broadening of SC transition and thus the induced disorder is, in a great extent, of
the ``granular'' type (with typical length scale of the order of $\xi$). It is in accordance
with the conception of random impurity potential introduced by excess oxygen atoms which may
cause carrier localization. Electrons are concentrated in wells near random potential minima
(``metallic'' regions) while insulating regions are located near the potential maxima.

A following sequence of regimes may be outlined as the amplitude of random potential $\gamma$
diminishes relative to electron Fermi energy $\varepsilon_F$:
\begin{enumerate}
  \item {\em Strong localization regime}, $\gamma\gg\varepsilon_F$ ($h\sigma_s/e^2\ll 1$). It is
  highly inhomogeneous system with small localization radius of electrons, $R_{loc} < \xi$.
  Superconductivity is absent.
  \item {\em A vicinity of insulator - ``metal'' transition}, $\gamma\cong\varepsilon_F$
  ($k_F\ell\cong 1$). It is still inhomogeneous system. Regions with $R_{loc} > \xi$ are formed,
  but they do not cover all the plane. Superconducting state with broad SC - transition develops.
  A coexistence of superconductivity and localization on the same $\rho(T)$ dependency is most
  probable just in this region.
  \item {\em ``Metallic'' (weak localization) regime}, $\gamma\ll\varepsilon_F$ ($k_F\ell\gg 1$).
  It is homogeneous, superconducting system with well defined SC-transition.
\end{enumerate}

Fig.5a shows the in-plane resistivity data for Nd$_{2-x}$Ce$_x$CuO$_{4+\delta}$ samples ($x =
0.15$) with different degree of disorder induced by ion irradiation of a sample. The
irradiation of the optimally reduced Nd$_{1.85}$Ce$_{0.15}$CuO$_4$ sample (with
$T_c^{onset}=T_0=22.5$\,K and $R$(30\,K)=$R_0=1\,\Omega$) was carried out in circular
accelerator by He$^+$ ions ($E = 1.2$\,MeV) step by step up to $\Phi = 10^{16}$\,cm$^{-2}$\
[19]. It will be seen that the insertion of radiation defects leads to substantial increase of
the resistivity, to destruction of superconductivity and also provokes a transition to
insulator-like low-temperature behaviour. The dependencies both of the critical
SC-temperatures and normal state resistance at $T=30$\,K on fluence $\Phi$ are presented in
Fig.5b. It is seen that $T_c(R_0)$ tends to zero at $\Phi_c \cong 1 \times
10^{15}$\,cm$^{-2}$. The degree of $T_c$ decreasing ($dT_c/d\Phi$) becomes especially
appreciable in the region of substantial resistivity increase. Such a correlation of $T_c$
degradation due to irradiation with the value of $R$ has been observed as well in hole-doped
cuprates [20, 21].

\section{Concluding remarks}

We have investigated temperature dependencies of the in-plane resistivity, $\rho_{ab}$, for
two series of Nd$_{2-x}$Ce$_x$CuO$_{4+\delta}$ single crystal films ($x=0.15$ and $x=0.18$)
for variable annealing conditions. In accordance with the data of the other authors [4-7] we
have observed that deoxygenation of a sample results in decrease of a value of resistivity as
a whole, i.e. a degree of disorder diminishes in process of reducing. The evolution of
low-temperature behaviour of $\rho_{ab}(T)$ from localization to superconductivity takes place
in tight correlation with the degree of disorder (estimated as a value of the parameter
$k_F\ell\equiv hc/e^2\rho_{ab}$).

It is important that SC-transition is observed only for samples with not high disorder,
$k_F\ell > 1$ ($k_F\ell\ge 2.5$ for $x=0.18$ and $k_F\ell\ge 10$ for $x=0.15$) as well as for
MBE films with $x=0.18$ [4] and for bulk single crystals with $x=0.15$ [5]. The appearance of
superconductivity just in a vicinity of transition from strong to weak localization (insulator
- ``metal'' transition) as well as the essential broadening of SC - transition in the process
of $T_c$ degradation may be described in the model of disorder two-dimensional system with
random impurity potential induced by nonstoichiometric oxygen defects.


\section*{Acknowledgements} This work was supported by the \hspace{0.5mm} RFBR \hspace{0.5mm}
grant No.~00-02-17427, State Contract No.~107-1(00)-P, and by Competition of Ural
Branch of RAS for young scientists, grant No. 10.


\newpage

Table. \vspace{8mm}

\begin{tabular}{|c|c|c|c|} \hline
  Sample & $T_{min},K$          & $\rho_{ab}(T_{min})$, m$\Omega$cm & $k_F\ell$ \\ \hline
  1      & 125                  & 2.3                               & 0.7      \\ \hline
  2      & 100                  & 0.148                             & 10.5     \\ \hline
  3      & 50                   & 0.068                             & 22.8     \\ \hline
  4      & $T_c^{onset}=19$     & 0.051                             & 30.4     \\ \hline
  5      & $T_c^{onset}=23.5$   & 0.0283                            & 54.8     \\ \hline
\end{tabular}

\newpage

\begin{center}
Captions to figures
\end{center}

Fig.1~(a) Temperature dependencies of the in-plane resistivity for three \linebreak
Nd$_{1.82}$Ce$_{0.18}$CuO$_{4+\delta}$ samples with different degree of disorder. The values
of parameter $k_F\ell$ are: 1 - 0.25; 2 - 2.5; 3 - 25;

(b)~The relative resistivity values as a function of temperature for the samples 1 and 3. The
inset shows the low-temperature data for sample 2. The dashed line shows the data in magnetic
field $B=1.3$\,T

\vspace{5mm}

Fig.2 The in-plane resistivity of Nd$_{1.82}$Ce$_{0.18}$CuO$_{4+\delta}$ sample 2 as a
function of temperature in different magnetic fields parallel (a) and perpendicular (b) to
$ab$-plane

\vspace{5mm}

Fig.3~(a) Temperature dependencies of the in-plane resistivity for
Nd$_{1.85}$Ce$_{0.15}$CuO$_{4+\delta}$ samples with different degree of disorder;

(b)~The relative resistivity values as a function of temperature for the same samples. The
dashed line shows the data in magnetic field $B=5.5$T

\vspace{5mm}

Fig.4~(a) The low-temperature data of the in-plane resistivity for
Nd$_{1.85}$Ce$_{0.15}$CuO$_{4+\delta}$ sample 2. The inset shows the data in perpendicular to
$ab$-plane magnetic fields $B$ up to 5.5~T;

(b)~The temperatures of the onset of SC-transition and of its completion as functions of
disorder parameter $k_F\ell$ of Nd$_{1.85}$Ce$_{0.15}$CuO$_{4+\delta}$ samples 2 - 5

\vspace{5mm}

Fig.5~(a) Temperature dependencies of the in-plane resistivity for
Nd$_{1.85}$Ce$_{0.15}$CuO$_{4+\delta}$ single crystal films at different irradiation fluence
$\Phi$(cm$^{-2}$): $\Phi=0$ (curve 1); $10^{14}$(2); $5\times 10^{14}$(3); $10^{15}$(4) and
$2\times 10^{15}$(5);

(b)~The relative values of resistance and the SC-transition temperature as functions of the
irradiation fluence for Nd$_{1.85}$Ce$_{0.15}$CuO$_{4+\delta}$ film


\begin{thebibliography}{99}

\bibitem{iv1} E. Dagotto, Rev. Mod. Phys. {\bf 66}, 763 (1994).


\bibitem{iv2}
V.\,V. Kapaev, Yu.\,V. Kopaev, XXI Symposium on Low Temperature Physics (Moskow 1998); A.\,I.
Ponomarev, A.\,N. Ignatenkov, L.\,D. Sabirzyanova et al., Proc. Int. Conf. Phys. Semicond.
(Ierusalem 1998), CD-ROM, Sec.5, Subsec.A, No.36.

\bibitem{iv10} A.\,A. Ivanov, S.\,G. Galkin, A.\,V. Kuznetsov et al.,
Physica C {\bf 180}, 69 (1991).

\bibitem{iv3} S. Tanda, S. Ohzeki, T. Nakayama, Phys. Rev. Lett. {\bf 69}, 530 (1992).

\bibitem{iv4} T.~Fujita, N.~Kikugawa, M.~Ito et al.,
Physica~C {\bf 341-348}, 1937 (2000).

\bibitem{iv5} N.\,A.~Fortune, K.~Murata, Y.~Yokoyama et al.,
{\em Physica~C} {\bf 178}, 437 (1991).

\bibitem{iv6} W.~Jiang, J.\,L.~Peng, Z.\,Y.~Li, R.\,L.~Greene,
Phys. Rev. B {\bf 47}, 8151 (1993).

\bibitem{iv7} J.\,I.~Martin, A.~Serguis, F.~Prado et al.,
Physica~C {\bf 341-348}, 1943 (2000).

\bibitem{iv8} P.\,A.~Lee and T.\,V.~Ramakrishnan,
Rev. Mod. Phys. {\bf 57}, 293 (1985).

\bibitem{iv9} G.\,I.~Harus, A.\,N.~Ignatenkov, A.\,I.~Ponomarev et al.,
JETP Lett. {\bf 70}, 97 (1999).

\bibitem{iv11} X.\,Q.~Xu, S.\,N.~Mao, Wu~Jiang et al.,
Phys. Rev. B {\bf 53}, 871 (1996).

\bibitem{iv12} M.\,V.~Sadovskii ``Superconductivity and Localization'',
World Scienific Publishing Co. Pte. Ltd., Singapore (2000).

\bibitem{iv13} L.\,N.~Bulaevskii, M.\,V.~Sadovskii,
JETP Lett. {\bf 39}, 640 (1984); J. Low-Temp. Phys. {\bf 59}, 89 (1985).

\bibitem{iv14} A.\,N.~Ignatenkov, A.\,I.~Ponomarev, L.\,D.~Sabirzyanova et al.,
JETP {\bf 92}, 1084 (2001).

\bibitem{iv15} D.~Belitz, T.\,R.~Kirkpatrick,
Rev. Mod. Phys. {\bf 66}, 261 (1994).

\bibitem{iv16} A.\,E.~White, R.\,C.~Dynes and J.\,P.~Garno,
Phys. Rev. B {\bf 33}, 3549 (1986).

\bibitem{iv17} J.\,M.~Valles, R.\,C.~Dynes and J.\,P.~Garno,
Phys. Rev. B {\bf 40}, 6680 (1989).

\bibitem{iv18} Y.~Fukuzumi, K.~Mizuhashi, K.~Takenaka and S.~Uchida,
Phys. Rev. Lett. {\bf 76}, 684 (1996).

\bibitem{iv19}
V.\,F.~Elesin, A.\,A.~Ivanov, I.\,A.~Rudnev et al., Superconductivity {\bf 5}, 514 (1992).

\bibitem{iv20} S.\,V. Atonenko, V.\,V. Evsstigneev, V.\,F. Elesin et al., Pis'ma v ZhETF {\bf
46}, 362 (1987).

\bibitem{iv21} G.\,J. Clark, A.\,D. Marwik, R.\,H. Koch and R.\,B. Labowitz,
Appl.~Phys.~Lett. {\bf 51}, 139 (1987).

\end{thebibliography}
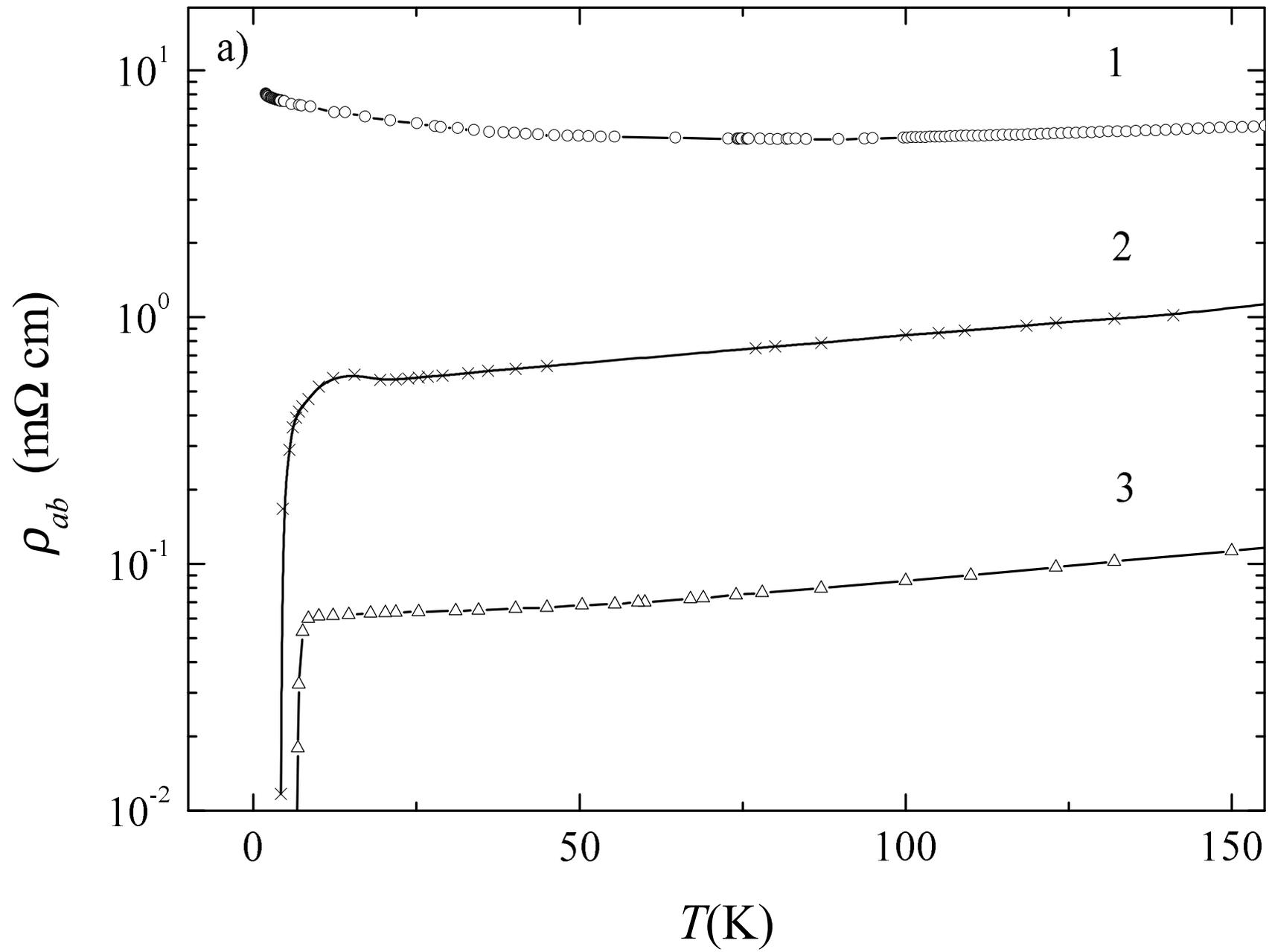
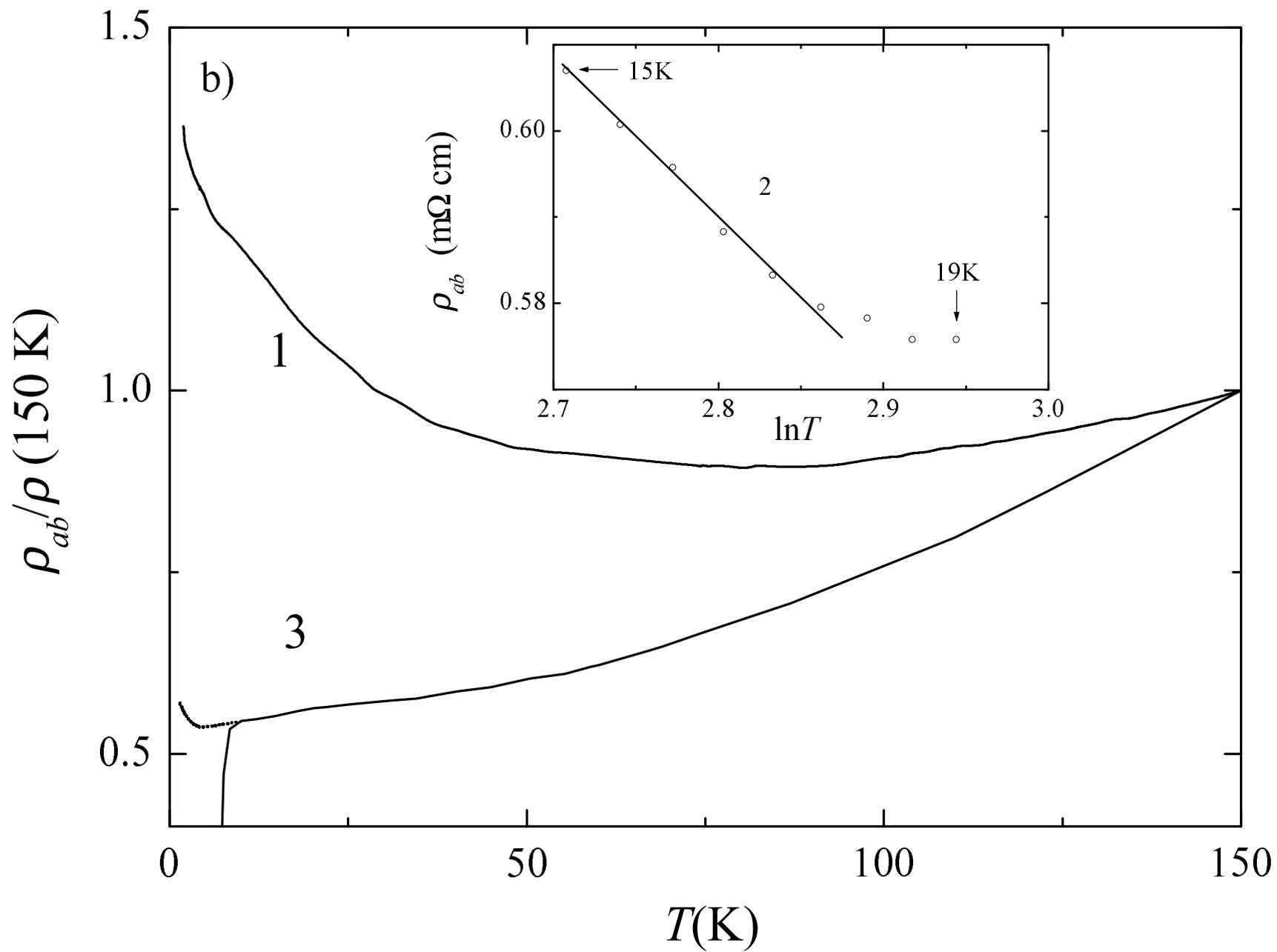
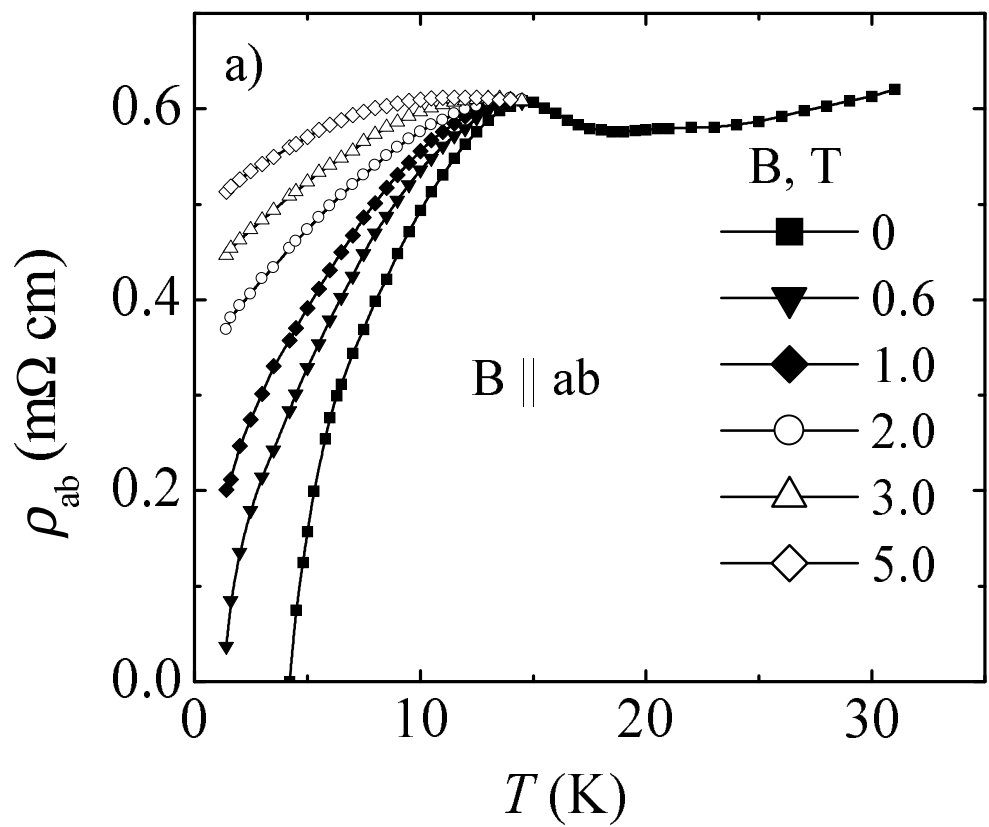
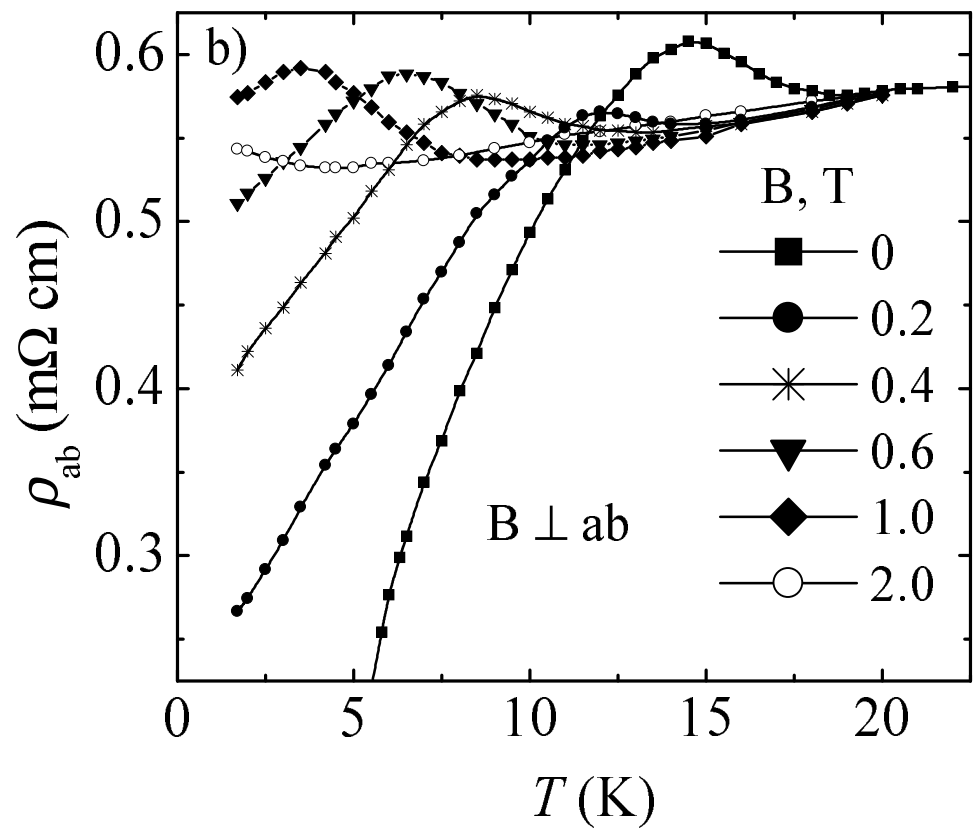
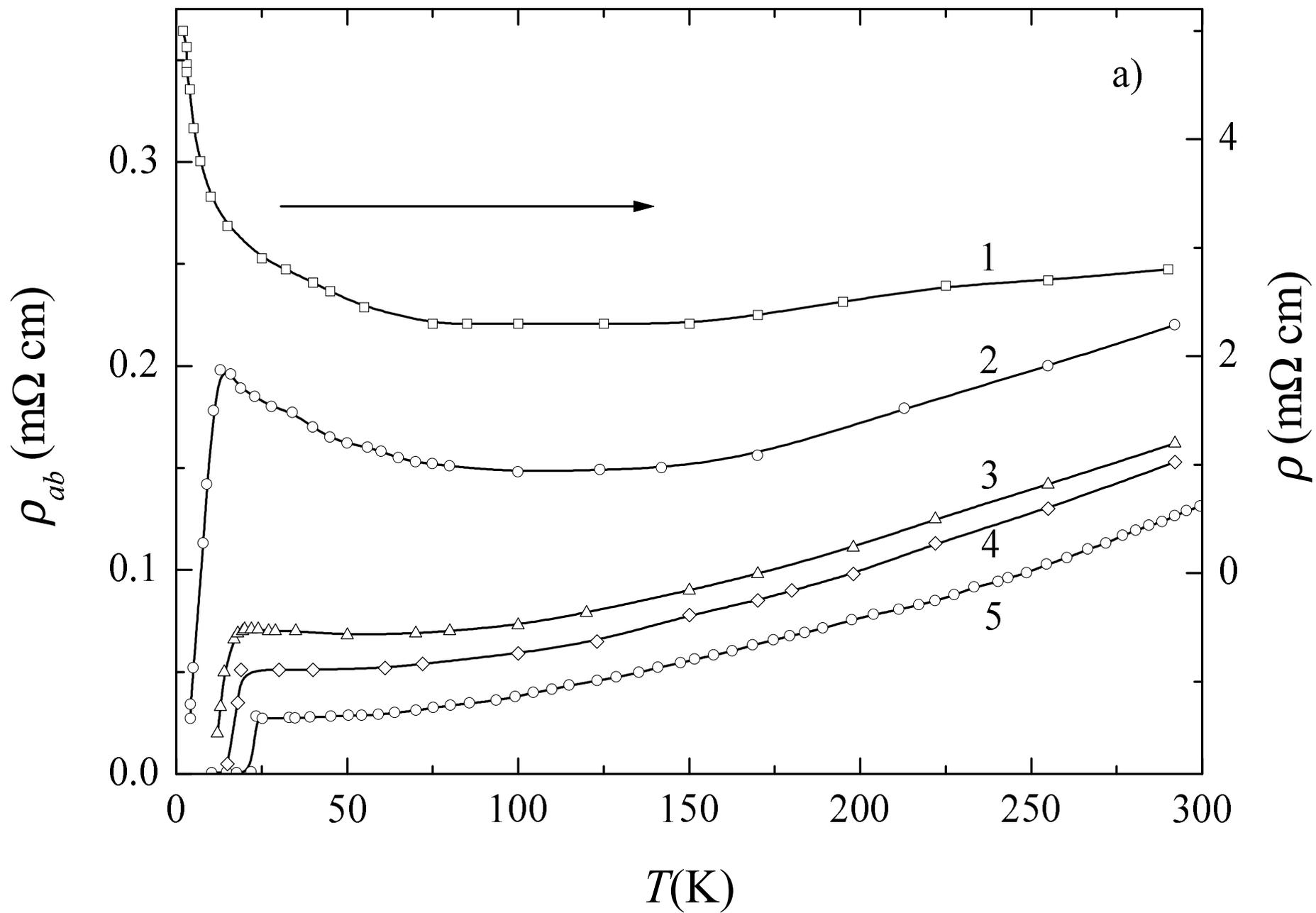
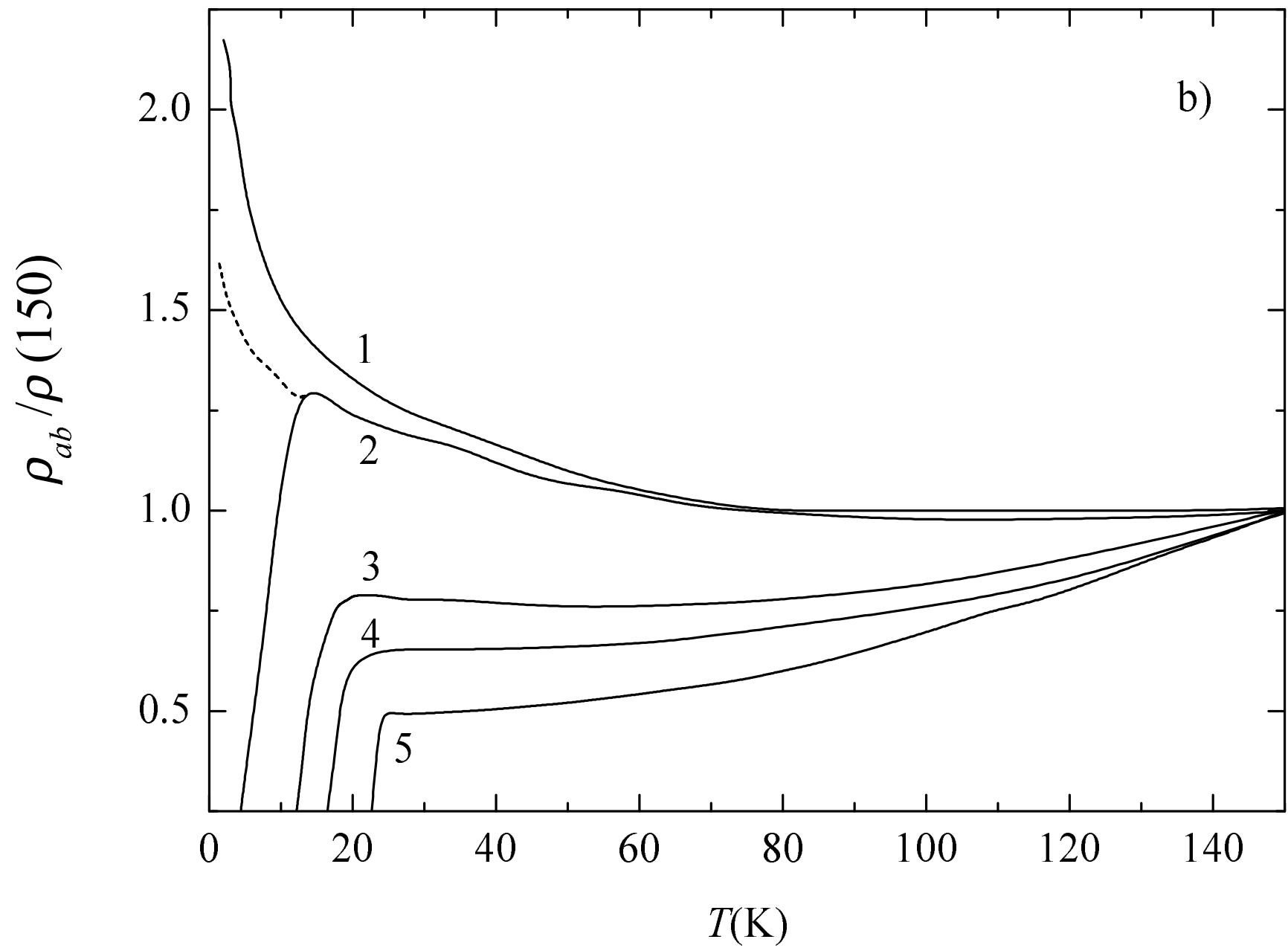
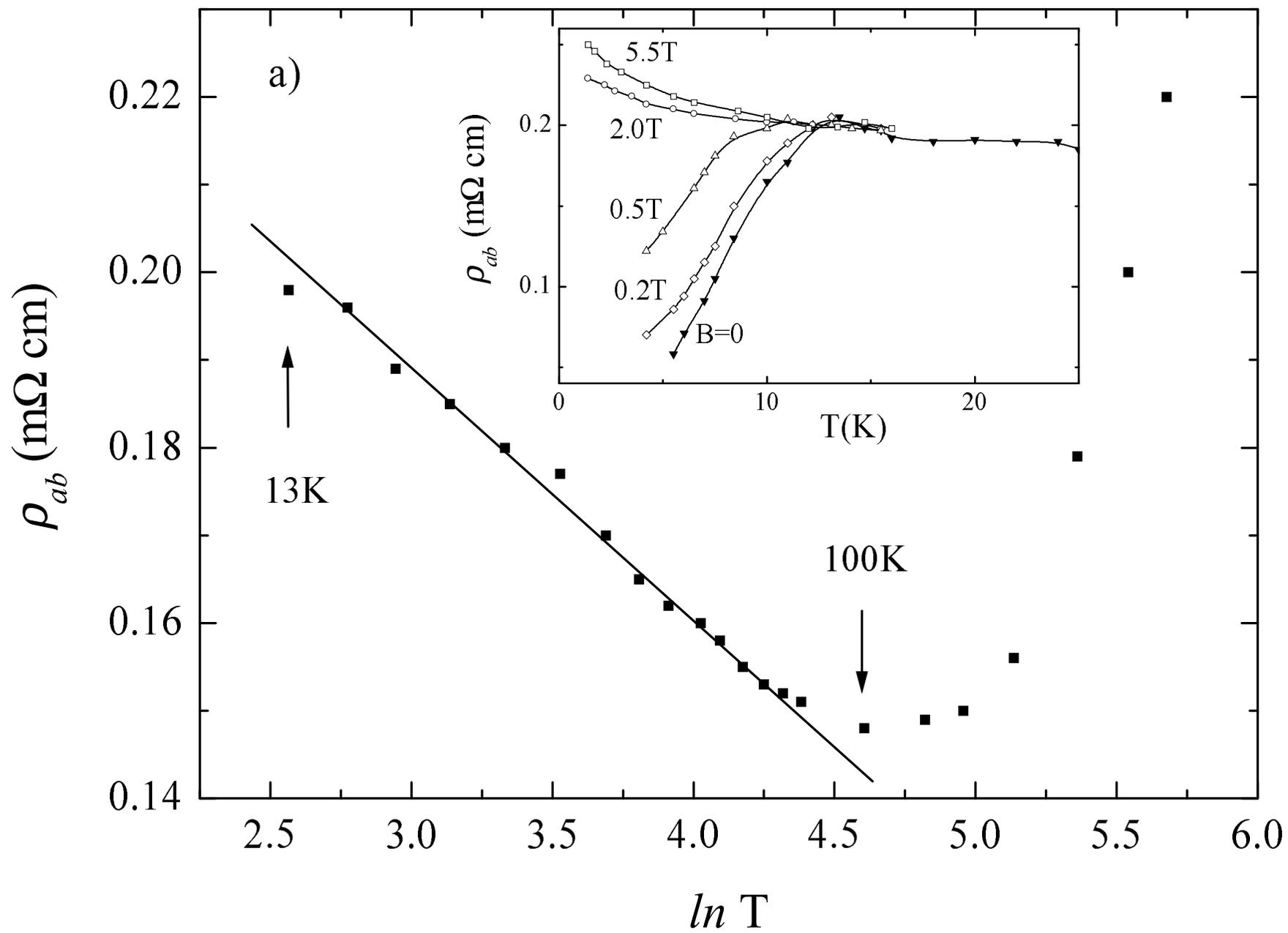
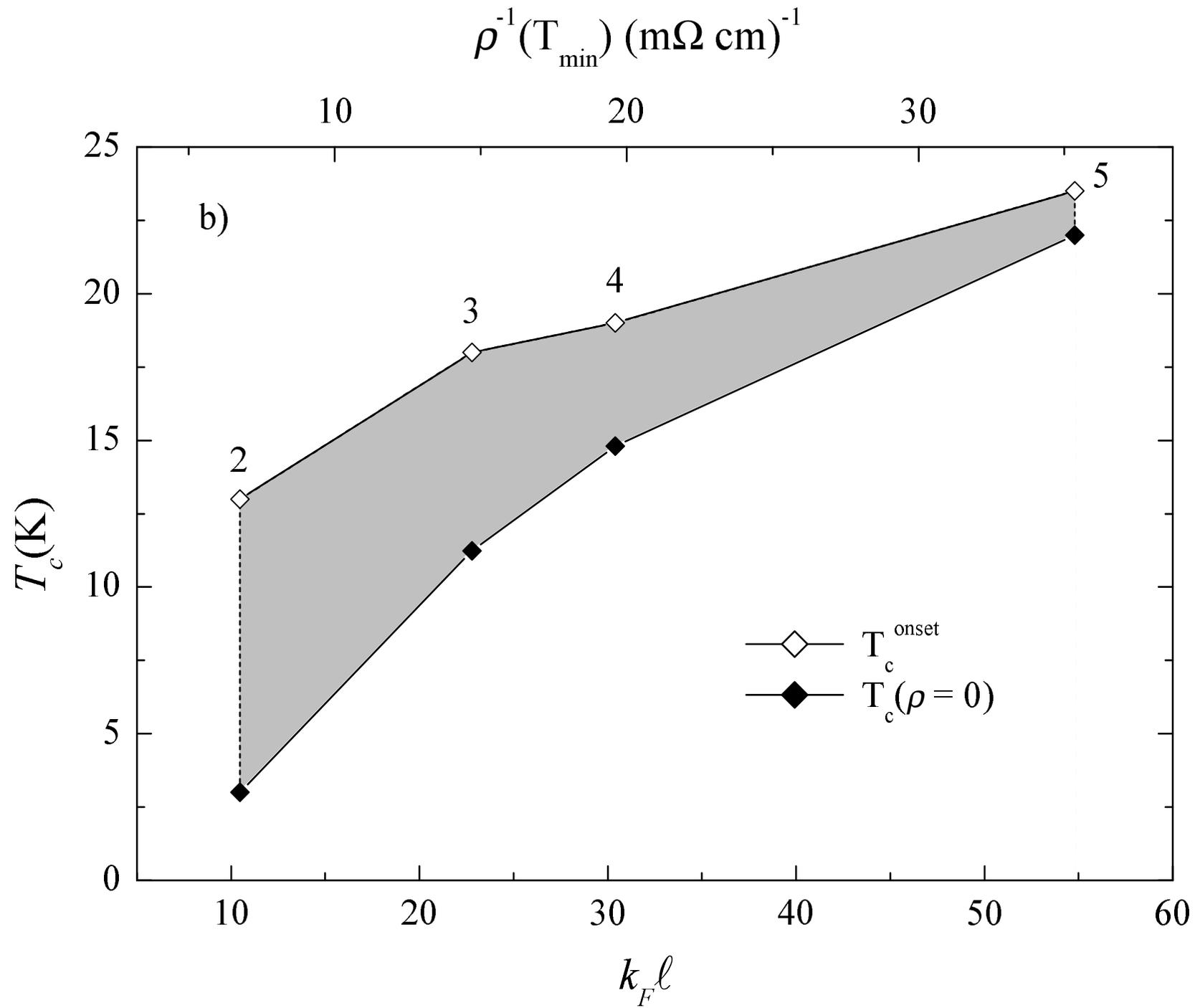
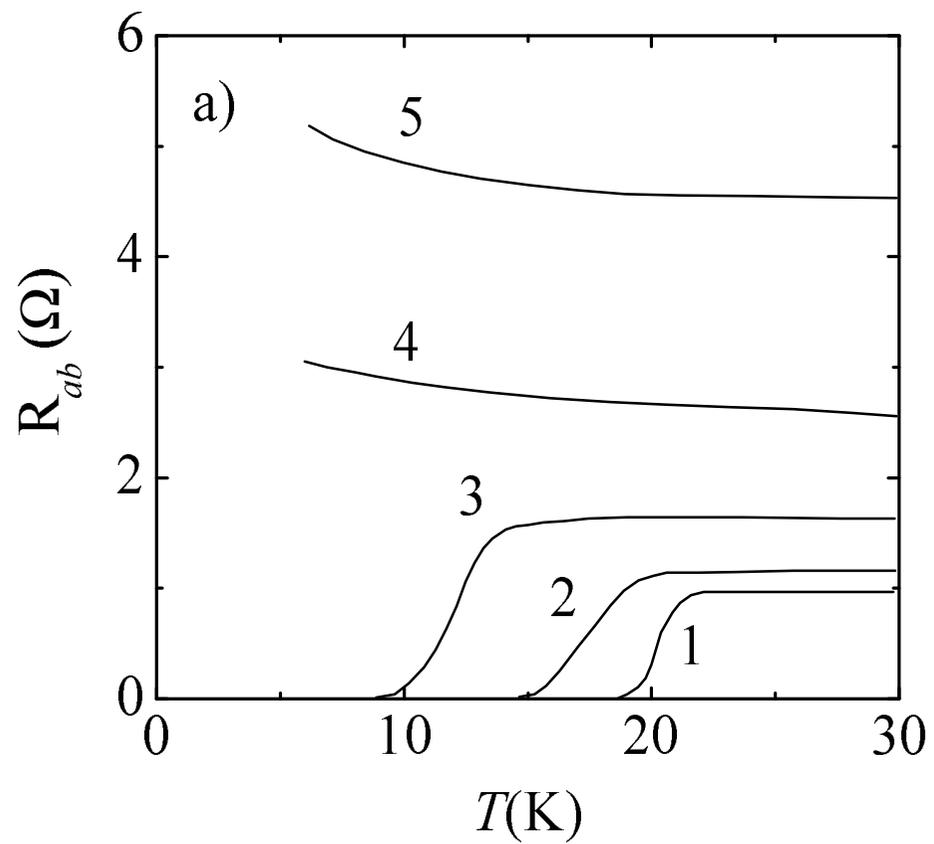
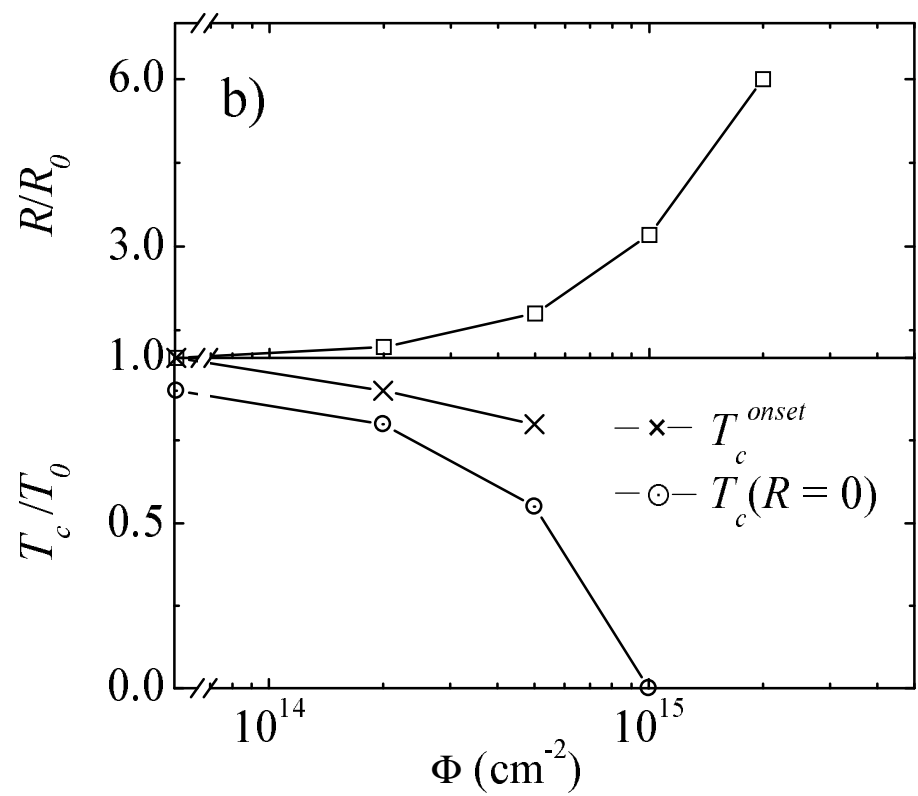
\end{document}